\documentclass[aps,prb,twocolumn,superscriptaddress,floatfix,nofootinbib]{revtex4-1}
\usepackage[utf8]{inputenc}
\usepackage[dvipsnames]{xcolor}
\usepackage{graphicx}
\usepackage{amsmath}
\usepackage{multirow}
\usepackage{comment}
\usepackage{url}
\usepackage{pifont}

\usepackage{colortbl}

\usepackage{filecontents}

\usepackage{xr}
\makeatletter
\newcommand*{\addFileDependency}[1]{
  \typeout{(#1)}
  \@addtofilelist{#1}
  \IfFileExists{#1}{}{\typeout{No file #1.}}
}
\makeatother

\newcommand*{\myexternaldocument}[1]{%
    \externaldocument{#1}%
    \addFileDependency{#1.tex}%
    \addFileDependency{#1.aux}%
}
\myexternaldocument{si}

\begin{document}

\title{Factors influencing quantum evaporation of helium from polar semiconductors from first principles}

\author{Lakshay Dheer}
\affiliation{Materials Sciences Division, Lawrence Berkeley National Laboratory, Berkeley, California 94720, USA}
\affiliation{Molecular Foundry, Lawrence Berkeley National Laboratory, Berkeley, California 94720, USA}
\author{Liang Z. Tan}
\affiliation{Molecular Foundry, Lawrence Berkeley National Laboratory, Berkeley, California 94720, USA}
\author{S. A. Lyon}
\affiliation{Department of Electrical Engineering, Princeton University, Princeton, New Jersey 08544, USA}
\author{Thomas Schenkel}
\affiliation{Accelerator Technology and Applied Physics Division, Lawrence Berkeley National Laboratory, Berkeley, California 94720, USA}
\author{Sin\'ead M. Griffin}
\altaffiliation{Contact for correspondence, SGriffin@lbl.gov}
\affiliation{Materials Sciences Division, Lawrence Berkeley National Laboratory, Berkeley, California 94720, USA}
\affiliation{Molecular Foundry, Lawrence Berkeley National Laboratory, Berkeley, California 94720, USA}

\date{\today}

\begin{abstract}\noindent While there is much indirect evidence for the existence of dark matter (DM), to date it has evaded detection. Current efforts focus on DM  masses over $\sim$GeV -- to push the sensitivity of DM searches to lower masses, new DM targets and detection schemes are needed. In this work, we focus on the latter - a novel detection scheme recently proposed to detect ~10-100 meV phonons in polar target materials. Previous work showed that well-motivated models of DM can interact with polar semiconductors to produce an athermal population of phonons. This new sensing scheme proposes that these phonons then facilitate quantum evaporation of $^3$He from a van der Waals film deposited on the target material. However, a fundamental understanding of the underlying process is still unclear, with several uncertainties related to the precise rate of evaporation and how it can be controlled. In this work, we use \textit{ab initio} density functional theory (DFT) calculations to compare the adsorption energies of helium atoms on a polar target material, sodium iodide (NaI), to understand the underlying evaporation physics. We explore the role of surface termination, monolayer coverage and elemental species on the rate of He evaporation from the target material. Using this, we discuss the optimal target features for He-evaporation experiments and their range of tunability through chemical and physical modifications such as applied field and surface termination.

\end{abstract}

\maketitle


\section{Introduction}
\label{sec:intro}
As the direct detection of dark matter pushes into the sub-GeV range, new ideas that extend the sensitivity of detection schemes are needed\cite{Essig_et_al:2012}. One such approach is the creation and subsequent detection of athermal quasiparticle excitations such as phonons through their interactions with dark matter\cite{Essig_et_al:2016,Knapen_2018_polar_materials, Griffin_GaAs_Sapphire, anthony-petersen_applying_2023}  . However, for the detection of such low-energy rare events remains challenging, and is typically addressed by state-of-the-art sensors kinetic-inductance detectors (KIDs)~\cite{Zmuidzinas:2012}, metallic magnetic calorimeters~\cite{Fleischmann_et_al:2005}, and transition-edge sensors (TES)~\cite{Irwin_et_al:1995,Romani_et_al:2018}. These detection schemes rely on the transport of the quasiparticle to the sensor and its efficient transport across the interface into the sensor, significantly reducing their efficiencies\cite{Harrelson_et_al:2021}.

Recent work, proposed to instead use quantum evaporation of He atoms from semiconductor targets as a single phonon detection scheme\cite{lyon_single_2024}. In this approach, athermal phonons cause the evaporation of 3-He from a van-der-Waals bonded layer on the target's surface. These evaporated He are then detected by the interaction of the nuclear spin and the spin state in a entangled pair of electron spin qubits. Because of this, this sensing scheme has potential sensitivity down to the single phonon regime, making it competitive with other phonon-based sensing schemes for low mass DM detection. Similar arguments also apply for studies of Coherent Elastic Neutrino Nucleus scattering, where detectors with sensitivity to (single) athermal phonons can extend reach towards tests of proposed beyond standard model neutrino physics~\cite{huber_snowmass_2022}, as well as applications in nuclear reactor monitoring and safeguards~\cite{akimov_observation_2017,choi_exploring_2023,  ackermann_final_2024} .

However, at these energy scales the details of the surface energetics become relevant including the surface termination, morphology and overlayer coverage. Moreover, having a route to selectively control the adsorption of He ions through materials selection and design would allow for a greater range of tunability for single-phonon detection. In this work we perform a systematic study of He adsorption on a polar semiconductor, NaI, which has previously been proposed as a promising low-mass DM detector target for phonon production\cite{Griffin_GaAs_Sapphire}. We consider the interaction of He on various crystallographic terminations of NaI and uncover underlying mechanisms of He evaporation. Using \textit{ab initio} DFT calculations we examine the effect of surface termination, monolayer coverage, and external electric fields on He-adsorption from NaI.

\section{Methods}

\begin{figure}
\includegraphics[width=0.4\textwidth]{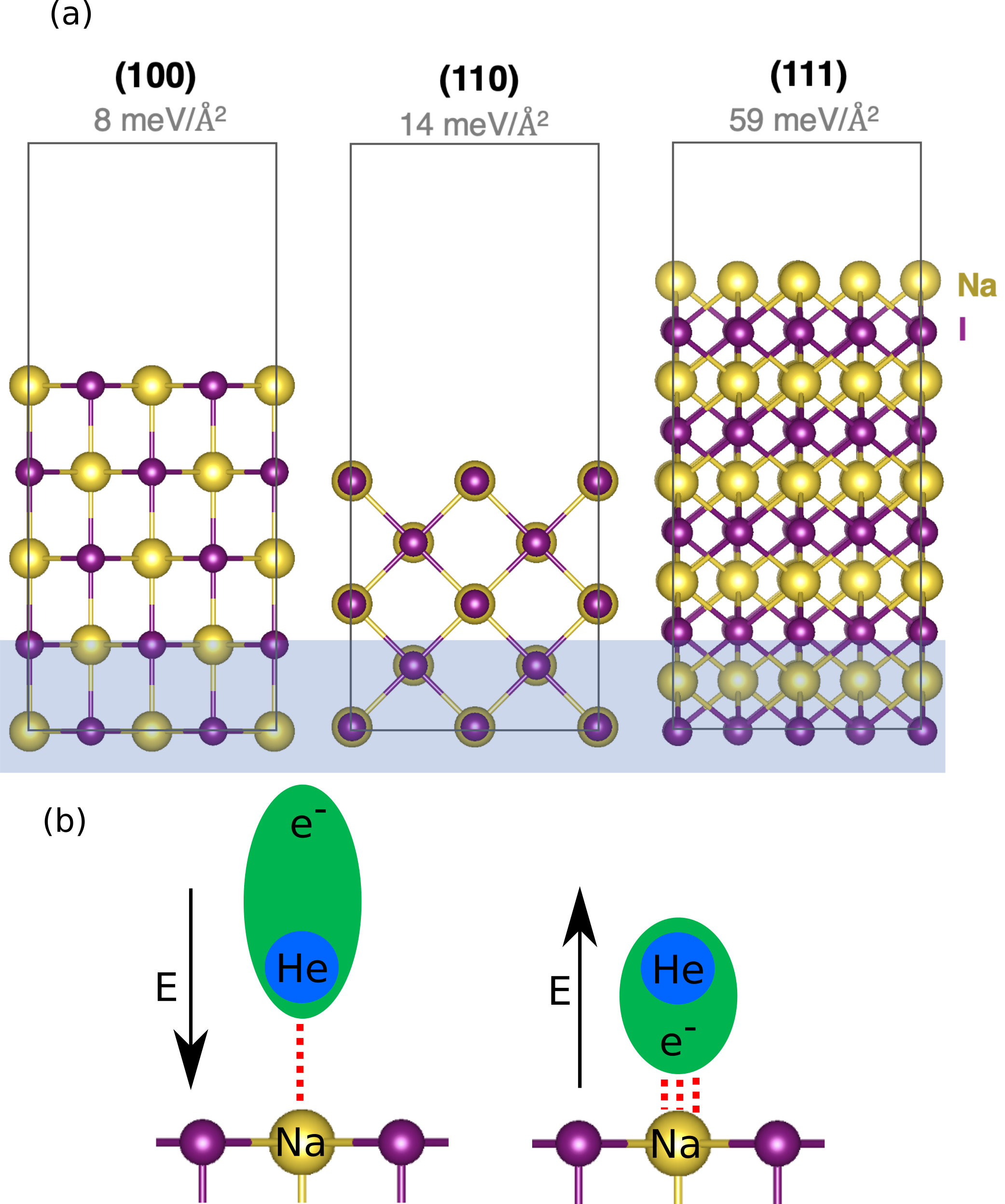}
\caption{(a) Three surface terminations of NaI considered for this study: (100), (110) and (111) where both the Na- and I-terminated (111) surfaces were calculated. The resulting surface energy of each termination is given with (100) being the most stable. The shaded region indicates those atoms that were fixed during cell optimizations. (b) Schematic illustrating the effect of external electric field ($E$) on the deformation of the electron density (green) and the strength of van der Waals forces (red). } 
\label{surface-terms}
\end{figure}

Our first-principles calculations are based on density functional theory (DFT) as implemented in the Quantum ESPRESSO code~\cite{giannozzi_quantum_2009} employing a plane-wave basis set and PAW pseudopotentials to represent the interaction between ionic cores and valence electrons. We adopt the exchange-correlation energy functional of Perdew-Burke-Ernzerhof (PBE) within a generalized gradient approximation (GGA), and occupation numbers of electronic states have been smeared with Fermi–Dirac distribution and a smearing width ($k_B$T) of 0.04 eV. Additionally, van der Waals (vdW) interactions using the Grimme scheme have been included to account for dispersive interactions. A kinetic energy cutoff of 65 Ry on the plane-wave basis is used for the Kohn-Sham wave functions, and a cutoff of 520 Ry is used for the charge density. All structures are optimized through the minimization of energy until the Hellmann-Feynman forces on each atom are smaller than 0.03 eV/Å in magnitude. A supercell approach is used to model the various surface terminations of NaI by introducing a vacuum layer of 20 Å thickness parallel to the slab separating adjacent periodic images. Each supercell contains a slab of 5 atomic planes, with each plane of the surfaces containing 2x2 NaI units. The surfaces studied in this work are shown in Fig.\ref{surface-terms}, with their respective surface energies also shown. Brillouin-zone integrations were sampled on a uniform grid of 6x6x1 k-points. 

\section{Results}

\begin{figure*}
\includegraphics[width=0.95\textwidth]{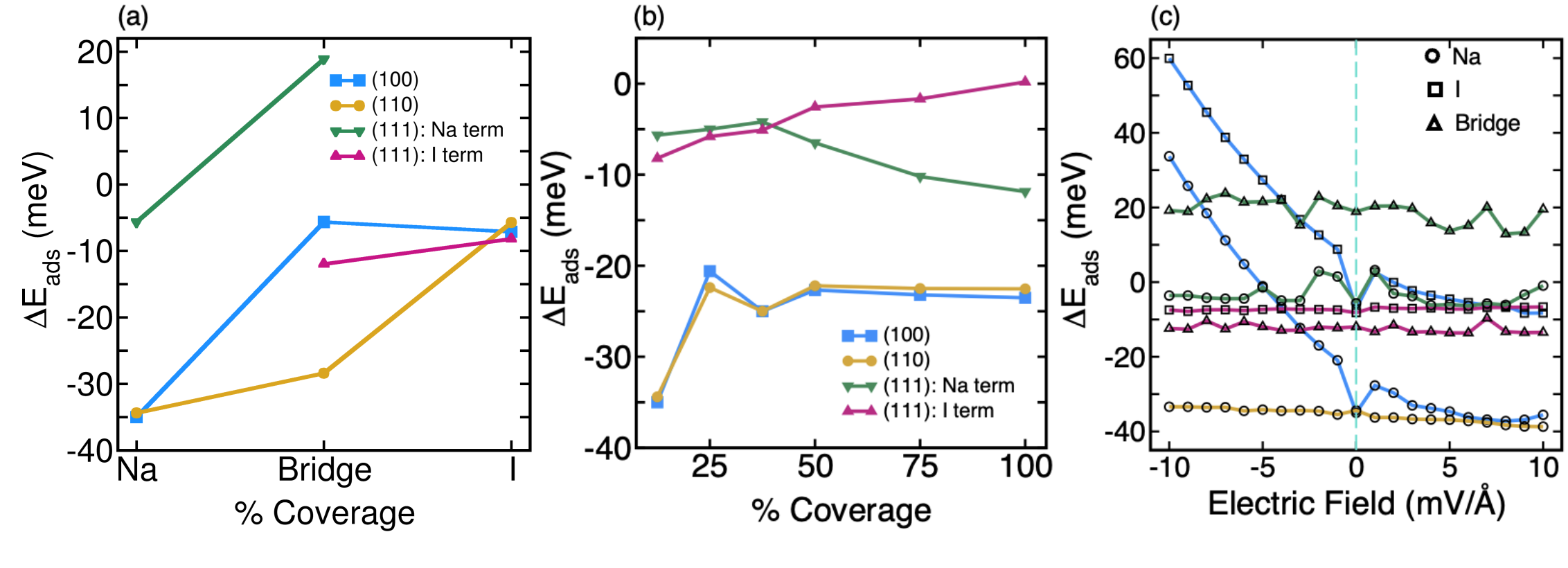}
\caption{Calculated adsorption energy of He for the (100), (110) and (111) surface terminations with varying (a) inequivalent adsorption sites Na, I and bridge, (b) surface coverage of monolayer He and (c) applied electric field.} 
\label{combined}
\end{figure*}

Sodium iodide (NaI) adopts the rock salt structure (space group \textit{Fm$\overline{3}m$}), with sodium and iodine forming two interpenetrating face-centered cubic lattices. From X-ray diffraction measurements, the three most thermodynamically stable crystallographic facets: (100), (110), and (111) were chosen for this work. Each layer of the (100) and (110) surfaces comprises of an equal number of Na and I atoms and thus they are non-polar. The (111) surface has an inherent electric field as it consists of alternating layers of Na and I atoms. Both Na and I terminations of the (111) surface were studied.

To assess the stability of the various surfaces of NaI, we calculated the surface energy (S.E.) with respect to the bulk NaI using:

\begin{equation}
    S.E.=\frac{E_{surface} - E_{bulk} * n}{2*S.A.}    
\label{eq_surface}
\end{equation}

where $n$ is the number of repeating units and S.A. is the surface area. For the (100), (110), and (111) surfaces calculated $S.E.$s are 8.03 meV/\AA$^2$, 14.20 meV/\AA$^2$ and 59.30 meV/\AA$^2$ respectively. To account for the intrinsic polarity in the (111) surface of NaI, $S.E.$ was also computed with dipole correction (59.8 meV/\AA$^2$) and the difference was found to be negligible when compared to the $S.E.$ without dipole correction. 

\subsection{Adsorption of He on NaI surfaces}
We considered the adsorption of He atoms on all inequivalent surface sites of NaI, namely, the Na site, I site, and bridge site. Note that the bridge site in the (100) and (110) surfaces is Na-I bond but for the (111) surface,Na-Na and I-I bond serve as the bridge site for the Na and I terminated surfaces, respectively. For the dilute He limit, i.e. one He atom per supercell, the adsorption energy (\textit{$\Delta$E$_{ads}$}) was calculated using:
\begin{equation}
    {\Delta}E_{ads} = E_{NaI,s+He}-(E_{NaI,s}+\mu_{He})
    \label{ads}
\end{equation}
  where, $E_{NaI,s+He}$, $E_{NaI,s}$, $\mu_{He}$ are the  energies of NaI surface-He complex, optimized NaI surface and the isolated He atom, respectively. A negative value of \textit{$\Delta$E$_{ads}$} denotes attractive interaction of He with the surface.

We first investigated the adsorption energy of He in the dilute limit for different surface terminations and adsorption sites. For all surface terminations we find that He exhibits a stronger adhesion to the Na site over I or the bridge site (see Fig.~\ref{combined}(a)). He shows very weak binding (\(\sim \) -10 meV) to the I site of all surfaces. Of the 4 surfaces, He showcases stronger binding to the Na site of the (100) and (110) surfaces and weaker adhesion to the Na site of the (111) surface. However, no such trend can be seen for \textit{$\Delta$E$_{ads}$} on bridge site since the adsorption sites differ here in the chemical environment depending on the surface morphology. We find a relatively stronger binding on the bridge site of the (110) surface however, on (111):Na term surface, He exhibits a repulsive interaction. For the (100) and (111):I terminated surfaces, He shows relatively weaker binding (\(\sim\) -10 meV) at the bridge site.

\begin{figure}
\includegraphics[width=0.49\textwidth]{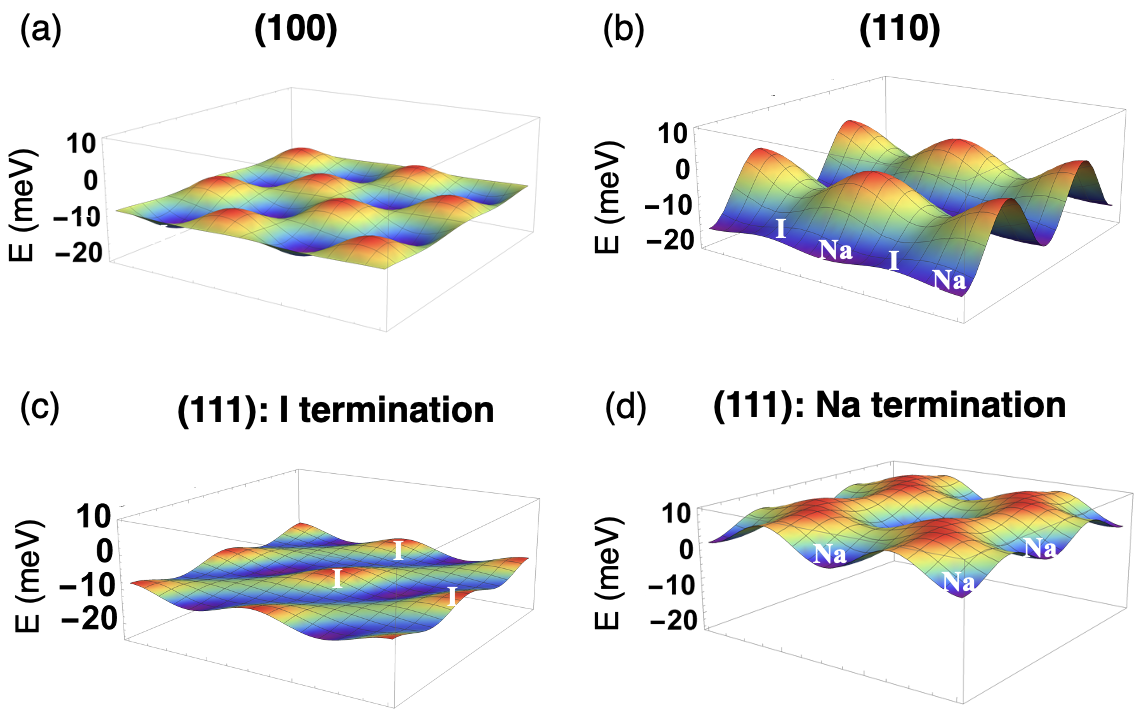}
\caption{Calculated potential energy surfaces for single He adsorption on various surface terminations of NaI as indicated. } 
\label{PES}
\end{figure}

We next calculated the potential energy surfaces (PES) of He interacting with the 4 surfaces studied here (Fig. \ref{PES}). The site-dependent energy of adsorption was calculated as a function of (x,y), keeping the z-coordinate fixed, where x and y are coordinates of position of He atom. We fitted the adsorption energy to a 2-dimenstional Fourier series to generate the functional form of the PESs. The data set comprised of 25, 20 and 30 adsorption sites for the (100), (110) and (111) surfaces, respectively. These PESs provide the landscape available to He on different NaI surfaces. We found Na and I sites to be the local minima and local maxima, respectively, regardless of the surface termination.

\subsection{Effect of He coverage}

\begin{figure}
\includegraphics[width=0.4\textwidth]{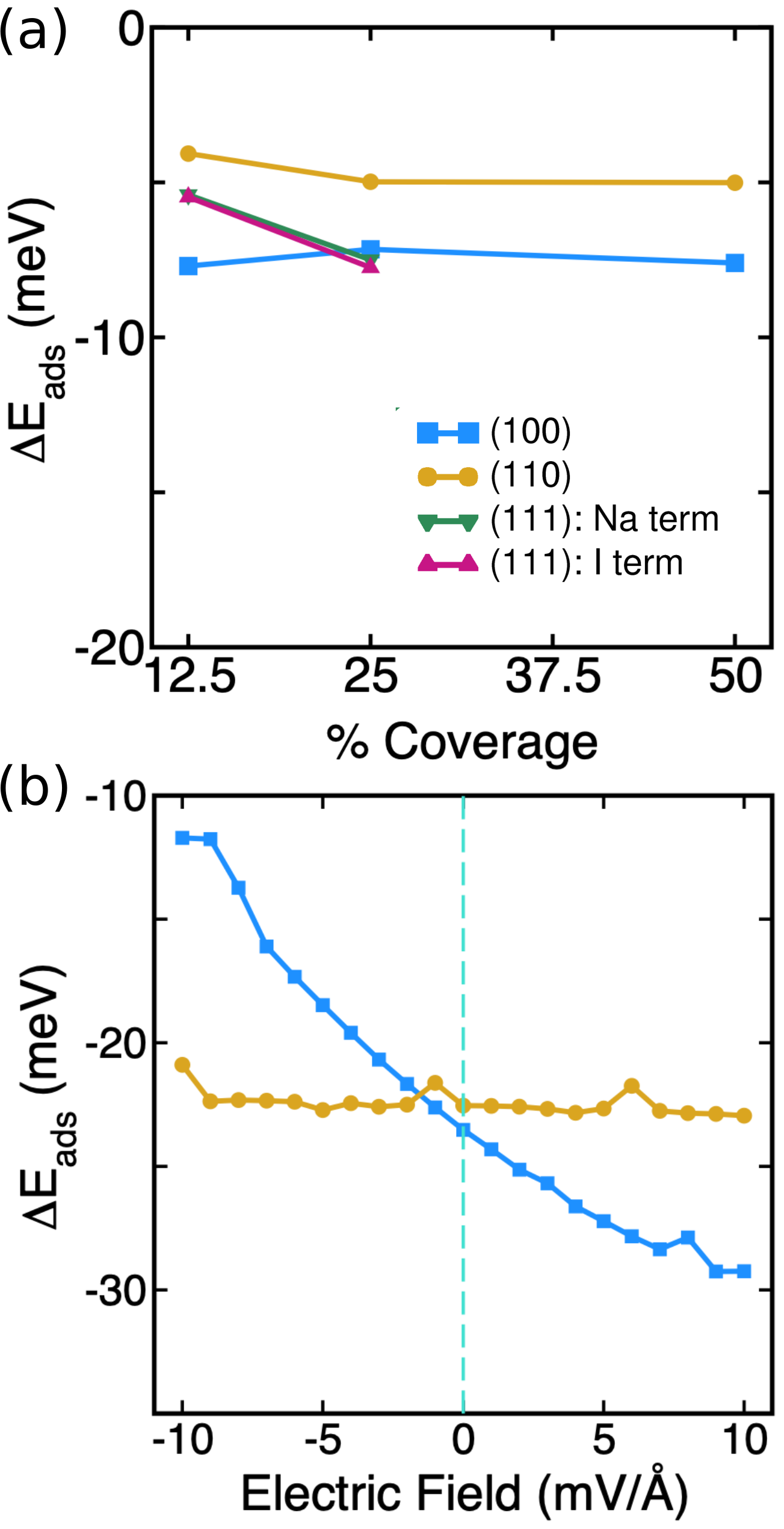}
\caption{(a) Calculated adsorption energy of He for increasing second-layer coverage for (100), (110) and (111) surface terminations. (b) Calculated adsorption energy of He for monolayer coverage as a function of applied electric field, for the (100) and (110) surface terminations. } 
\label{bilayer}
\end{figure}

While He showed the strongest adhesion to the Na site in the dilute limit -- regardless of surface termination -- we next investigate the influence of surface coverage on the resulting adsorption energies for different terminations. Our results are summarized in Fig.\ref{combined}(b). We find similar trends for the (100) and (110) surface terminations, namely the adsorption energy increases from the single-atom dilute limit ($\sim$-35 meV) to a steady value of ($\sim$-20-- -25 meV) over 25\% coverage owing to screening from the extra adsorbates. In contrast, we see that the adsorption energy steadily decreases for the (111) I terminated surface. However, for the (111) Na terminated surface, the adsorption energy remains steady at ~5 meV up to 37.5\% where it gradually increases up to ~12 meV.  We see a similar effect for bilayer coverage where the adsorption energy difference become independent of coverage beyond the dilute limit (Fig.\ref{bilayer}a).

\subsection{Applied Electric Fields}
Since the interaction between He and the semiconductor surfaces is governed by fluctuating dipoles that comprise van der Waals interactions, we next explored the use of applied electric fields for controlling He adsorption. Our results are summarized in Fig.~\ref{combined}(c). For each configuration considered (surface termination and adsoprtion site), an electric field was applied in direction parallel to the surface normal (that is, in the out-of-plane direction in Fig.\ref{surface-terms}). We find the strongest influence of applied electric field on the (100) surface with the adsorption energy going from 60 meV to -5 meV for I sites and 34 meV to -38 meV for Na sites with an applied field going from -10 mV/\AA\ to +10 mV/\AA. We find a much smaller change for the other terminations, with both the (111):Na terminated and (110) surfaces having a slight field-dependence of the adsorption energy, whereas the (111):I terminated remains constant for both adsorption sites.  We also investigated the electric-field-dependence of the adsorption energies for full monolayer surface coverage, finding similar trends to the dilute He case. We find that the adsorption energy changes from -12 meV to -29 meV for (100), whereas remains essentially constant for the (110) termination (Fig\ref{bilayer}b).

\section{Discussion}
Our first principles calculations find that several materials factors influence the adsorption energy of He on polar semiconductors, and will ultimately govern the efficiency of quantum evaporation from these systems. Firstly, we find that surface termination can result in the adsorption energy varying both in the nature (sign) of the interaction and its strength. We find here for the case of NaI that the adsorption energies in the dilute limit range between  -35 meV and +20 meV, depending on surface termination. We next examined how He coverage influences the adsorption, seeing a screening effect to come into play for larger coverage. Finally we examine how applied electric fields can change the adsoprtion energies, finding them to range between
 -60 meV and 52 meV for an applied field from -10 mV/\AA\ to +10 mV/\AA. This suggests that applying electric fields on the appropriately selected surfaces could provide a route to tunability of the quantum evaporation from He surfaces.

The observed trends of the adsorption energies can be understood as follows. The generally strong binding to Na sites is a result of the attractive interaction between the positively charged Na$^{+}$ ions and the deformable electron cloud of the He atoms. 
As He electron density is shifted away from the surface by the applied field, van der Waals forces between He electrons and the surface are weakened, leading to decreased adsorption. Reversing the direction of the field shifts the He electron cloud towards the surface, increasing adsorption (Fig.~\ref{surface-terms}b). However, it is easier to shift electrons away from the surface rather than shifting them to the surface because Coulombic repulsion from bulk NaI electrons limits the deformation of the electron density towards the surface, explaining the asymmetry of the electric field dependence of the adsorption energies in Fig.~\ref{combined}. These effects are the strongest on the 100 surface, especially when He electron density is shifted away from the surface. This is because the next-neighbor binding sites are weakly binding on the 100 surface as they are iodine sites, contributing little to binding to He. In contrast, the next-neighbor Na sites of the 110 and 111 surfaces contribute to binding even as He electrons are shifted away from the surface, leading to less dependence of adsorption on the applied electric fields.

Surface roughness and variations in morphology are expected to modify adsorption energies from the theoretical values computed here. For instance, an unequal distribution of He across the surface, or the presence of surface defects may affect the electronic screening of the local environment, leading to a distribution of adsorption energies on actual surfaces. Nevertheless, our results suggest that the electric field tunability of adsorption energies is likely to be a robust feature of the 100 surface, having shown significant electric field-dependence at the dilute and monolayer coverage limits. As the reason for this tunability arises from the bulk/vacuum asymmetry and the geometry of the 100 surface, we expect that it can only be affected by severe disruptions in the rock salt crystal structure at the surface. On actual surfaces with a distribution of He sites with different adsorption energies, electric fields can be applied to select only a subset of He sites to have adsorption energies in the few meV range suitable for phonon sensing.
In future work, electrical field tuning of the binding energies of other weakly bound adsorbates and polar crystal surfaces will be explored~\cite{ferm_infrared-induced_1987, maris_dark_2017} .  These can extend the available tuning range of surface binding energies for differential measurements of quantum evaporation with a series of phonon energy cut-offs and hence a range of dark matter candidates and neutral current neutrino scattering conditions.

\section{Acknowledgments}
This work was supported by the Quantum Information Science Enabled Discovery (QuantISED) for High Energy Physics (KA2401032). Work at the Molecular Foundry was supported by the Office of Science, Office of Basic Energy Sciences, of the U.S. Department of Energy under Contract No. DE-AC02-05CH11231.  This research used resources of the National Energy Research Scientific Computing Center (NERSC), a U.S. Department of Energy Office of Science User Facility operated under Contract No. DE-AC02-05CH11231.

\bibliographystyle{apsrev4-1} 
\bibliography{references}

\end{document}